\pdfoutput=1
\documentclass[aps,superscriptaddress,twocolumn,prd,nofootinbib,longbibliography]{revtex4-2}
\usepackage{amsmath, amssymb, amsfonts, amsthm, latexsym, epsfig, mathrsfs, xcolor, bbm, slashed,braket,bm}

\usepackage[inline]{enumitem}

\usepackage{setspace}

\usepackage[T1]{fontenc}
\usepackage[utf8]{inputenc}
\usepackage{lmodern}
\usepackage[normalem]{ulem}

\usepackage[unicode, colorlinks, allcolors=blue!70!black, linktocpage, pdfusetitle]{hyperref}
\usepackage[all]{hypcap}

\definecolor{indigo(dye)}{rgb}{0.0, 0.25, 0.42}

\usepackage{tikz}
\usetikzlibrary{arrows.meta}

\numberwithin{equation}{section}

\usepackage{cleveref}

\usepackage{cancel}

\usepackage{microtype}

\setlist[enumerate]{noitemsep, label=(\arabic*), ref=(\arabic*)}

\newlist{condlist}{enumerate}{2}
\setlist[condlist,1]{noitemsep, label=(\arabic*), ref=(\arabic*)}
\setlist[condlist,2]{noitemsep, label=(\alph*), ref=(\arabic{condlisti}.\alph*)}
\crefname{condlisti}{condition}{conditions}
\crefname{condlistii}{condition}{conditions}

\renewcommand\thesection{\arabic{section}}
\renewcommand\thesubsection{\arabic{subsection}}

\makeatletter
\def\p@subsection{\thesection.}
\def\p@subsubsection{\thesection.\thesubsection.}
\makeatother

\theoremstyle{plain}

\theoremstyle{definition}

\theoremstyle{remark}

\crefname{equation}{Eq.}{Eqs.}
\Crefname{equation}{Equation}{Equations}

\crefname{section}{Sec.}{Secs.}
\crefname{appendix}{Appendix}{Appendices}
\crefname{figure}{Fig.}{Figs.}

\crefname{definition}{Def.}{Defs.}
\crefname{prop}{Prop.}{Props.}
\crefname{lemma}{Lemma}{Lemmas}
\crefname{corollary}{Cor.}{Cors.}
\crefname{thm}{Theorem}{Theorems}
\crefname{remark}{Remark}{Remarks}

\crefname{ass}{Assumptions}{Assumptions}
\crefname{property}{Properties}{Properties}

\newcommand{\be}{\begin{equation}}
\newcommand{\ee}{\end{equation}}

\newcommand{\mc}{\mathcal}

\newcommand{\ms}{\mathscr}

\newcommand{\Lie}{\pounds}
\newcommand{\hatLie}{\Lie\kern-0.25em\hat{\vphantom{\Lie{}}}\kern0.25em}

\let\oldint\int
\renewcommand{\int}{\oldint\limits}

\newcommand{\op}[1]{\boldsymbol{#1}}

\newcommand{\Hilb}{\mathscr{H}}

\newcommand{\Fock}{\mathscr{F}}

\newcommand{\EM}{\textrm{EM}}

\newcommand{\GR}{\textrm{GR}}

\newcommand{\cev}[1]{\reflectbox{\ensuremath{\vec{\reflectbox{\ensuremath{#1}}}}}}

\begin{document}

\title{Local Description of Decoherence of Quantum Superpositions by Black Holes and Other Bodies}
\author{Daine L. Danielson}\email{daine@uchicago.edu}
\affiliation{Enrico Fermi Institute and Department of Physics, The University of Chicago, 933 East 56th Street, Chicago, Illinois 60637, USA}
\author{Gautam Satishchandran}\email{gautam.satish@princeton.edu}
\affiliation{Princeton Gravity Initiative, Princeton University, Jadwin Hall, Washington Road, Princeton, New Jersey 08544, USA}
\author{Robert M. Wald}\email{rmwa@uchicago.edu}
\affiliation{Enrico Fermi Institute and Department of Physics, The University of Chicago, 933 East 56th Street, Chicago, Illinois 60637, USA}

\date{\today}

\begin{abstract}
\noindent It was previously shown that if an experimenter, Alice, puts a massive or charged body in a quantum spatial superposition, then the presence of a black hole (or more generally any Killing horizon) will eventually decohere the superposition \cite{Danielson:2022tdw,Danielson:2022sga,Gralla:2023oya}. This decoherence was identified as resulting from the radiation of soft photons/gravitons through the horizon, thus suggesting that the global structure of the spacetime is essential for describing the decoherence. In this paper, we show that the decoherence can alternatively be described in terms of the local two-point function of the quantum field within Alice's lab, without any direct reference to the horizon. From this point of view, the decoherence of Alice's superposition in the presence of a black hole arises from the extremely low frequency Hawking quanta present in Alice's lab. We explicitly calculate the decoherence occurring in Schwarzschild spacetime in the Unruh vacuum from the local viewpoint. We then use this viewpoint to elucidate (i) the differences in decoherence effects that would occur in Schwarzschild spacetime in the Boulware and Hartle-Hawking vacua; (ii) the difference in decoherence effects that would occur in Minkowski spacetime filled with a thermal bath as compared with Schwarzschild spacetime; (iii) the lack of decoherence in the spacetime of a static star even though the vacuum state outside the star is similar in many respects to the Boulware vacuum around a black hole; and (iv) the requirements on the degrees of freedom of a material body needed to produce a decoherence effect that mimics that of a black hole. 

\end{abstract}
\maketitle
\section{Introduction}
In quantum mechanics, any interaction of a system with an ``environment'' will typically result in decoherence of the system. This decoherence arises because the environment responds differently depending on the state of the quantum system and thereby becomes entangled with the quantum system. While, in principle, any local ``environmental influences'' (i.e. interaction with any degrees of freedom present within the lab) can be minimized by a sufficiently controlled experiment, the long-range gravitational fields of the superposition cannot be perfectly controlled. In principle, any quantum superposition of gravitational fields can be ``measured'' by an external  observer---or the environment---and may give rise to some degree of decoherence. As was already noted by Feynman in the 1950s \cite{Dewitt_2011,Zeh_2008}, key insights into the quantum nature of gravity can be gleaned by considering gedankenexperiments analyzing the entanglement and decoherence due to the gravitational field of a massive body. Indeed, such gedankenexperiments have been the basis of actual proposed tabletop experiments in quantum gravity to measure the gravitationally mediated entanglement of two quantum systems \cite{FORD_1982,Lindner_2005,Bahrami_2015,Bose_2017,Marletto_2017,Carney_2019, Haine_2021, Qvarfort_2020, Carlesso_2019, Howl_2021, Matsumura_2020, Pedernales_2021, Liu_2021, Datta_2021, Gonzalez-Ballestero:2021, Krisnanda:2020uh, doi:10.1126/sciadv.abg2879, CHRISTODOULOU201964, Bose:2022uxe,Gonzalez-Ballestero:2021}.

In previous work \cite{Danielson:2022tdw,Danielson:2022sga} (see also \cite{Gralla:2023oya}), we showed that a black hole can, in effect, measure the long range fields of a massive or charged body, resulting in the decoherence of a quantum superposition of such a body. The precise mechanism producing this decoherence was found to be entangling radiation that is emitted by the quantum superposition into the black hole. To understand this, suppose an experimenter, Alice, creates a spatial superposition of a charged or massive body, e.g., by putting it through a Stern-Gerlach apparatus. Suppose that after keeping this superposition in place for a time, $T$, Alice brings the components of the body together and determines if they have remained coherent. Even if Alice performs her experiment in Minkowski spacetime, some entangling radiation will be emitted to infinity when the spatial superposition is created and brought back together. However, in Minkowski spacetime, the decoherence resulting from this radiation can be made arbitrarily small by ``opening'' and ``closing'' the superposition in a sufficiently adiabatic manner. Furthermore, in Minkowski spacetime, the amount of time, $T$, that she keeps the superposition open is not relevant to the decoherence (provided, of course, that she makes all ordinary interactions with the environment negligible). However, as we showed, this is not the case in the presence of a black hole. Although the energy radiated into the black hole can be made arbitrarily small by ``opening'' and ``closing'' the superposition is a sufficiently adiabatic manner, the number of entangling photons/gravitons radiated into the black hole increases linearly with total time $T$ that the superposition is kept ``open,'' so, eventually, a black hole will decohere any quantum superposition. This effect occurs more generally for any Killing horizon, e.g., it also occurs for a Rindler horizon and a cosmological horizon \cite{Danielson:2022sga,Gralla:2023oya}.

The analysis of \cite{Danielson:2022tdw,Danielson:2022sga,Gralla:2023oya} strongly suggests that global aspects of the structure of the spacetime---specifically, the presence of a horizon---are essential for the decoherence effect. The main purpose of the present paper is to show that one can give an alternative, purely local description of the decoherence in terms of the behavior of the quantum field within Alice's lab. From this viewpoint, the decoherence arises from the behavior of the unperturbed two-point function of the quantum field in the region where the superposition was created. In particular, the decoherence in the presence of a black hole can be understood as resulting from the extremely low frequency Hawking radiation that partially penetrates into Alice's lab before being reflected back into the black hole by the effective potential of the black hole. This local viewpoint will enable us to gain insights into various aspects of the decoherence process, such as the differences in decoherence that occur in different vacuum states and in different spacetimes. We will also gain insight into the requirements on a material body to mimic the decoherence effects of a black hole. 

We note that, very recently, Wilson-Gerow, Dugad, and Chen \cite{Wilson-Gerow:2024ljx} also have given a local formulation of our decoherence results, focusing particularly on the Rindler case, i.e., an accelerating observer in Minkowski spacetime. The methods and arguments used in \cite{Wilson-Gerow:2024ljx} are quite different from the ones we shall give in our analysis below. Nevertheless, there are a number of significant points of overlap in the results. In particular, our result Eq.~(\ref{N2pt}) relating the decoherence to the local two-point function of the electric field corresponds to Eq.~(103) of \cite{Wilson-Gerow:2024ljx}.

We also note that in a previous paper \cite{Danielson:2022sga} we analyzed the decohering effects of the scattering of Unruh radiation on a charged superposition in an accelerating laboratory in Minkowski spacetime. We concluded that this decoherence was distinct from (and smaller than) the decohering effects of emission of entangling radiation through the Rindler horizon. However, in \cite{Danielson:2022sga} we considered only incoherent scattering effects of individual Unruh photons. We did not consider the coherent effects of the presence of a large number of Unruh photons of frequency $\omega \sim 1/T \ll 1/a$, where $a$ denotes the acceleration of the laboratory. As we shall see in the present paper, the presence of these very low frequency photons can be viewed as stimulating the emission of entangling radiation from the superposition. Thus, the decoherence effect in Rindler spacetime is, in fact, intimately related to the presence of very low frequency Unruh radiation in the Minkowski vacuum. Similarly, the decoherence effect in a black hole spacetime is intimately related to the presence of very low frequency Hawking radiation in the Unruh vacuum.

Our local reformulation of the decoherence makes manifest that one can interpret the decoherence of Alice's superposition in terms of the interaction of Alice's particle with the local state of the quantum field in her lab. It should be emphasized that the thermal nature of the state is, by itself, insufficient to account for this effect~\cite{Danielson:2022sga,Wilson-Gerow:2024ljx}. In particular, for the decoherence in the Unruh vacuum in the presence of a black hole, it is essential that there is an extremely large reservoir of ``soft'' Hawking quanta in the Unruh vacuum as compared with an ordinary inertial thermal bath in Minkowski spacetime at the same temperature. Furthermore, in the Boulware vacuum in a black hole spacetime---which is the ground state with respect to the timelike Killing field and thus has no particles---Alice's superposition still spontaneously emits entangling soft photons/gravitons into the black hole, but the number of entangling particles grows only logarithmically with time. The Unruh vacuum corresponds to a thermal population whose density of states diverges at low frequencies. The presence of these low-frequency quanta stimulate the emission of entangling soft radiation into the horizon, so that the number of entangling soft photons/gravitons grows linearly in time. 

Our local reformulation of Alice's decoherence also allows one to also consider what happens when one replaces the black hole by a body without a horizon. It is instructive to consider the case where Alice's lab is in the spacetime outside of a static, spherical star rather than a black hole but we do not consider any internal degrees of freedom of the matter composing the star, i.e., we consider only the effect of replacing the spacetime geometry of a black hole with the spacetime geometry of a star. If the quantum field is in its stationary ground state in the spacetime of the star, then the two-point function of the quantum field in Alice's lab should look very much like the Boulware vacuum in Schwarzschild spacetime with respect to the incoming modes from infinity. However, the ``white hole incoming modes'' of Schwarzschild will be entirely absent for the star. These white hole modes are responsible for the decoherence effects that grow with $T$ in Schwarzschild, so a similar decoherence will not occur for the star. Even if the quantum field is in a thermal state in the spacetime of the static star, there will be no decoherence effects that grow with $T$. Thus, the presence of a horizon is essential for the kind of decoherence obtained for a Schwarzschild black hole.

Nevertheless, one can get decoherence without a horizon if one has a material body with internal degrees of freedom that interact electromagnetically and/or gravitationally with the particle in Alice's lab. In this situation, the interaction is now mediated by the long-range Couloumbic/Newtonian field of the superposition without any emission of radiation, analogous to the gedankenexperiment \cite{Belenchia_2018, Danielson:2021} in flat spacetime where Alice and Bob both perform their experiments adiabatically and in causal contact with one another.
As we shall show, the material body will mimic the decoherence effects of the black hole if, at very low frequencies, the thermal fluctuations of its electric dipole moment and/or mass quadrupole moment agree with black hole case [see Eqs.~(\ref{fluctdip}) and (\ref{fluctquad}) below]. 
This issue has recently been investigated by Biggs and Maldacena \cite{Biggs:2024dgp}. In order for a body of size comparable to that of a black hole to be able to absorb and emit low frequency electromagnetic or gravitational waves as efficiently as the black hole, a conducting or gravitating body must have a very large resistance or viscosity. There does not appear to be any difficulty, in principle, in achieving this in the electromagnetic case \cite{Biggs:2024dgp}. However, some extraordinary physical properties of matter would be required to mimic the quantum gravitational decoherence effect \cite{Biggs:2024dgp}.

In Sec.~\ref{sec:gedanken}, we review our previous derivation of decoherence in the presence of a horizon. In Sec.~\ref{sec:local}, we provide a local reformulation of this decoherence in terms of the two-point function of the quantum field in Alice's laboratory over the duration of her experiment. In Sec. \ref{locsch}, we compute the decoherence in the Unruh vacuum in Schwarzschild using our local formulation, which requires the computation of the two-point function of the electric field along the worldline of Alice's lab. Finally, in Sec. \ref{sec:comp}, we compute the decoherence for different vacua in Schwarschild and in different spacetimes, including a brief discussion of the decoherence due to entanglement with an ordinary material body. 

Unless otherwise stated, we will work in Planck units where $G=c=\hbar=k_{\textrm{B}}=1$ and, in electromagnetic formulas, we also put $\epsilon_0 = 1$. We will generally follow the notational conventions of \cite{Wald:1984rg}. In particular, abstract spacetime indices will be denoted with lowercase latin indices from the early alphabet ($a,b,c \dots$). Spacetime coordinate components will be denoted with Greek indices. Spatial coordinates and components will be denoted with Latin indices from the middle alphabet ($i,j,k,\dots$).

\section{Decoherence of a Quantum Superposition Due to Radiation}
\label{sec:gedanken}
In this section we briefly review the analysis of decoherence due to radiation through a Killing horizon previously given in \cite{Danielson:2022tdw,Danielson:2022sga}. We will focus on the electromagnetic case and merely state the corresponding results in the gravitational case.

An experimenter, Alice, in a stationary lab in a stationary spacetime $(\mathscr M, g_{ab})$ controls a charged particle\footnote{The ``particle'' need not be ``elementary,'' e.g., it could be a nanoparticle. All that is required is that the degrees of freedom of the particle apart from its center of mass may be neglected.} which is initially held stationary in her lab. The particle is put through a Stern-Gerlach apparatus over a time $T_{1}$ so that  at the end of this process its quantum state is of the form 
\begin{equation}
\ket{\psi} = \frac{1}{\sqrt{2}}(\ket{\psi_{1}} + \ket{\psi_{2}})
\end{equation}
where $\ket{\psi_{1}}$ and $\ket{\psi_{2}}$ are the spatially separated, normalized states of the particle after passing through the Stern-Gerlach apparatus. Alice maintains this stationary superposition for a (proper) time $T$, and she subsequently recombines her particle over a time $T_{2}$ where we assume that $T\gg T_{1},T_{2}$. The recombined particle is then kept stationary. We now analyze the decoherence of Alice's particle due to emission of entangling electromagnetic radiation sourced by Alice's superposition. 

We assume that $\ket{\psi_{1}}$ and $\ket{\psi_{2}}$ are sufficiently spatially separated that $\braket{\psi_{1}|\op{j}^{a}|\psi_{2}} = 0$ and we further assume that the fluctuations of the charge current $\op{j}^{a}$ in states $\ket{\psi_{1}}$ and $\ket{\psi_{2}}$ are negligible compared with their expected values. We may then treat the charge-current of each component of the superposition as a $c$-number source in Maxwell's equations. Thus, if Alice's particle is in state $\ket{\psi_{n}}$ for $n=1,2$, then the electromagnetic field operator is given by \cite{Yang:1950vi}
\begin{equation}
\label{eq:gensol}
\op{A}_{n,a} = \op{A}_{a}^{\textrm{in}} + G^{\textrm{ret}}_{a}(j_{n})\op{1}
\end{equation}
where $\op{A}_{a}^{\textrm{in}}$ is the unperturbed (``in'') field operator and $G^{\textrm{ret}}_{a}(j_{n})$ is the retarded solution associated to the classical charge-current $j_{n}^a = \braket{\psi_{n}|\op{j}^{a}|\psi_{n}}$. The ``out'' radiative field at late times is obtained by subtracting the final Coulomb field $C_a$ of the recombined particle from $\op{A}_{n,a}$
\begin{eqnarray}
\label{eq:out}
\op{A}^{\textrm{out}}_{n,a} &=& \op{A}_{n,a} - C_{a} \op{1} \nonumber \\
&=& \op{A}^{\textrm{in}}_{n,a} + \mathcal{A}_{n,a}\op{1}
\end{eqnarray}
where
\begin{equation}
\label{Aidef}
\mathcal{A}_{n,a}\equiv G^{\textrm{ret}}_{a}(j_{n}) - C_{a} \, .
\end{equation}

We assume that the initial state of the quantum electromagnetic field is some ``vacuum state'' (i.e., a pure, quasifree state) $\ket{\Omega}$ that is invariant under the time translation symmetries of the spacetime. The unperturbed field operator $\op{A}^{\textrm{in}}$ on the Fock space, $\Fock(\Hilb_{\textrm{in}})$, associated with $\ket{\Omega}$
can be expressed in terms of annihilation and creation operators on $\Fock(\Hilb_{\textrm{in}})$ as 
\begin{equation}
\label{eq:Af}
\op{A}^{\textrm{in}}_{a}(f^{a}) = i \op{a}(\overline{K\Delta(f)}) - i \op{a}^{\dagger}(K\Delta(f))
\end{equation}
where $f^{a}$ is a divergence-free\footnote{Restriction of the smearing to divergence-free test functions is
necessary and sufficient to eliminate the gauge dependence of $\op{A}_{\textrm{in},a}$
(see, e.g., p. 101 of \cite{Wald_1995}).} test vector field and $\Delta(f)$ is the advanced minus retarded solution to Maxwell's equation with source $f^a$
\begin{equation}
[\Delta(f)]_a(x) = \int_{\mathscr M} \sqrt{-g}d^{4}x^{\prime}\Delta_{aa^{\prime}}(x,x^{\prime})f^{a^{\prime}}(x^{\prime})
\end{equation}
where $\Delta_{aa^{\prime}}(x,x^{\prime})$ is the advanced minus retarded Greens function. Here $K$ is the map that takes classical solutions into the corresponding one-particle states in the Fock space defined by $\ket{\Omega}$.

As can be seen from Eq.~(\ref{eq:out}), the ``out'' state corresponding to the ``in'' vacuum $\ket{\Omega}$ has field correlation functions at late times that are obtained from the vacuum correlation functions by shifting the field operator by a multiple of the identity operator. It follows that if Alice's particle is in state $\ket{\psi_n}$, then the ``out'' state of the electromagnetic field will be given by the coherent state
\begin{equation}
\label{cohst}
\ket{\Psi_{n}}=e^{-\frac{1}{2}\lVert K\mathcal{A}_{n}\rVert^2}\exp\big[\op{a}^{\dagger}(K \mathcal{A}_{n})\big]\ket{\Omega}
\end{equation}
where, for notational simplicity, we drop the spacetime index ``$a$'' from $\mathcal{A}_{n,a}$, Eq.~(\ref{Aidef}), here and elsewhere in the remainder of this section. The norm $||K \mathcal{A}_n||$ appearing in Eq.~(\ref{cohst}) is taken in the one-particle Hilbert space of the Fock space of $\ket{\Omega}$.

The joint quantum state of Alice's particle together with the emitted electromagnetic radiation at late times is given by 
\begin{equation}
\frac{1}{\sqrt{2}} \big(\ket{\psi_{1}}\otimes \ket{\Psi_{1}} + \ket{\psi_{2}}\otimes \ket{\Psi_{2}}\big). 
\end{equation}
Thus, the decoherence of Alice's particle due to the emission of electromagnetic radiation is then given by 
\begin{equation}
\label{eq:decAlice}
\ms{D}_{\textrm{Alice}} = 1 - |\braket{\Psi_{1}|\Psi_{2}}| \, .
\end{equation}
The magnitude of the inner product of the coherent states $\ket{\Psi_1}$ and $\ket{\Psi_2}$ is computed to be
\begin{equation}
\label{eq:inprodcoh}
|\braket{\Psi_{1}|\Psi_{2}}| = \exp \bigg(-\frac{1}{2}||K(\mc{A}_{1}-\mc{A}_{2})||^{2}\bigg)
\end{equation}
where $K(\mc{A}_{1}-\mc{A}_{2})$ denotes the one-particle state associated with late time classical solution
\begin{equation}
\mc{A}_{1}-\mc{A}_{2} = G^{\textrm{ret}}(j_{1}- j_2).
\label{gret}
\end{equation}
But $||K(\mc{A}_{1}-\mc{A}_{2})||^{2}$ is equal to the expected number of photons, $\braket{N}$, in the coherent state associated with the late time classical solution $\mc{A}_{1}-\mc{A}_{2}$ sourced by $j_{1}- j_2$
\begin{equation}
\label{eq:entphot}
\braket{N}\equiv  ||K(\mc{A}_{1}-\mc{A}_{2})||^{2} = ||K G^{\textrm{ret}}(j_{1}- j_2)||^2 .
\end{equation}
Thus, we have
\begin{equation}
\label{eq:decAlice2}
\ms{D} = 1 - \exp\bigg(-\frac{1}{2} \braket{N}\bigg) \, .
\end{equation}
We shall refer to $\braket{N}$ as the expected number of {\em entangling photons}. If the expected number of entangling photons is significantly bigger than $1$, then Alice's superposition will be completely decohered.

Thus, we see that to compute the decoherence of a superposition created by Alice under the assumptions stated above, we proceed as follows: 

\begin{enumerate}

\item We compute the expected currents $j_1$ and $j_2$ of the components of Alice's superposition. 

\item We compute the classical retarded solution $G^{\textrm{ret}}(j_{1}- j_2)$ sourced by the difference of these currents. 

\item We compute the one-particle state $K G^{\textrm{ret}}(j_{1}- j_2)$ of this classical solution at late times and its squared norm $\lVert K G^{\textrm{ret}}(j_{1}- j_2)\rVert^2$. This yields the expected number of entangling photons, $\braket{N}$, and thereby the decoherence, Eq.~(\ref{eq:decAlice2}). Note that the one-particle map $K$ depends on the choice of vacuum state $\ket{\Omega}$.

\end{enumerate}

The above analysis extends directly to the linearized quantum gravitational case, where the linearized metric perturbation $h_{ab}$ is treated as a field propagating on a fixed spacetime background. In the above formulas, we simply replace $A_{a}$ with $h_{ab}$ and we replace the current $j_{a}$ with the linearized stress tensor $T_{ab}$. The expected number of entangling gravitons is then given by the analog of Eq.~(\ref{eq:entphot}) and the decoherence is given by Eq.~(\ref{eq:decAlice2}).

In Minkowski spacetime, we may take the notion of stationarity to be given by ordinary, inertial time translations and we may take $\ket{\Omega}$ to be the Poincaré invariant vacuum. If a particle of charge $q$ is put in a superposition separated by a distance $d$, then we may estimate $G^{\textrm{ret}}(j_{1}- j_2)$ near null infinity using the Larmor formula. The one-particle state $K G^{\textrm{ret}}(j_{1}- j_2)$ is the positive frequency part of this solution with respect to inertial time translations. The norm of this one-particle state is given by the Klein-Gordon norm. The expected number of entangling photons is thereby estimated to be \cite{Belenchia_2018,Danielson:2021} 
\begin{equation}
\label{eq:NEMMink}
\braket{N} \sim \frac{q^{2}d^{2}}{\textrm{min}[T_{1},T_{2}]^{2}} \quad \textrm{ (Minkowski, EM)}.
\end{equation}
Thus, the decoherence does not depend upon $T$ and can be made arbitrarily small by performing the separation and recombination of the superposition sufficiently slowly, so that $T_{1},T_{2}\gg qd$.

In the analysis of the corresponding gravitational case we must take into account the fact that conservation of total stress-energy implies that the center of mass cannot change. Thus if the component $\ket{\psi_1}$ of the superposition corresponds to the particle moving to the right, then Alice's lab must move a tiny bit to the left to keep the center of mass unchanged. The upshot is that the leading order contribution to the retarded solution with source $T^{ab}_{1} - T^{ab}_{2}$ arises from quadrupole radiation rather than dipole radiation. The estimate corresponding to Eq.~(\ref{eq:NEMMink}) for the number of entangling gravitons is \cite{Belenchia_2018,Danielson:2021}
\begin{equation}
\label{eq:NGRMink}
\braket{N} \sim \frac{m^{2}d^{4}}{\textrm{min}[T_{1},T_{2}]^{4}} \quad \textrm{ (Minkowski, GR)}.
\end{equation}
Again, the decoherence does not depend upon $T$ and can be made arbitrarily small by performing the separation and recombination of the superposition sufficiently slowly, so that $T_{1},T_{2}\gg \sqrt{md^{2}}$.

However, it was shown in \cite{Danielson:2022tdw} that the situation is drastically different in the presence of a black hole or, more generally, any Killing horizon \cite{Danielson:2022sga}. In the case of a black hole, the relevant vacuum is the ``Unruh vacuum'' $\ket{\Omega_{\textrm{U}}}$. If $T_1, T_2$ are sufficiently large---i.e., if Alice separates and recombines the particle sufficiently slowly---then the number of entangling photons/gravitons emitted to infinity will again be negligible. However, if an initially stationary source is moved to a new position and held there forever, the retarded solution will exhibit a ``memory effect'' on the horizon \cite{Hawking:2016msc}. Consequently, it can be seen that if Alice were to keep her superposition open forever, an infinite number of soft entangling photons/gravitons would be emitted through the horizon, in close analogy with the infrared divergences at infinity that arise in scattering theory (see, e.g.,  \cite{asymp-quant,Ashtekar:2018lor,Prabhu_2022,Prabhu:2024lmg}). If Alice closes her superposition after time $T$, then the number of entangling photons radiated through the horizon will be finite but will grow linearly with $T$. 
In the electromagnetic case the number of photons grows as \cite{Danielson:2022tdw}
\begin{equation}
\label{eq:NEMBH}
\braket{N} \sim \frac{M^{3}q^{2}d^{2}}{D^{6}}T \quad \textrm{ (black hole, EM)}
\end{equation}
where $M$ is the mass of the black hole and $D$ is the proper distance of Alice's lab from the horizon (and, for simplicity, we have assumed that $D\gtrsim M$ so that, e.g., the redshift factor at Alice's lab is of order unity and can be absorbed in the ``$\sim$''). The analogous computation in the gravitational case\footnote{In the gravitational case, it will be necessary to have some additional stress-energy present to hold Alice's lab stationary and keep her particle components stationary. We neglect any effects of such additional stress-energy.} yields \cite{Danielson:2022tdw}
\begin{equation}
\label{eq:NGRBH}
\braket{N} \sim \frac{M^{5}m^{2}d^{4}}{D^{10}}T \quad \textrm{ (black hole, GR)}.
\end{equation}
More generally, it was shown that in the presence of any Killing horizon (e.g., a Rindler or cosmological horizon) the number of entangling soft photons and gravitons grows linearly in the time $T$ that the superposition is maintained~\cite{Danielson:2022sga}.

The above results were obtained by calculating the quantum state of the electromagnetic and linearized gravitational fields on the horizon 
associated with the retarded solution sourced by the components of Alice's superposition. The decoherence of Alice's particle was attributed to the emission of entangling photons/gravitons through the horizon. Thus, it might appear that the global properties of the spacetime---specifically, the presence of a horizon---are essential for the description of the decoherence phenomenon we have just given. However, we will now show that the decoherence can alternatively be described purely in terms of the local properties of the unperturbed quantum field within Alice's laboratory. This alternative viewpoint will enable us to compare decoherence phenomena in the presence of a black hole with decoherence phenomena occurring when no horizon is present.

\section{Local Reformulation of the Decoherence}
\label{sec:local}

As in the previous section, we first consider the electromagnetic case and then state the corresponding results in the gravitational case.

A local reformulation of the electromagnetic decoherence results of the previous section is obtained from the following simple observations: First, since $j_1 = j_2$ at late times, the retarded solution $G^{\textrm{ret}}(j_{1}- j_2)$ is equal to $- \Delta (j_{1}- j_2)$ at late times, where $\Delta = G^{\textrm{adv}} - G^{\textrm{ret}}$. Thus, we may replace $G^{\textrm{ret}}$ by $-\Delta$ in Eqs.~(\ref{gret}) and (\ref{eq:entphot}), and we no longer have to evaluate these quantities at late times. Second, we note that it follows immediately from Eq.~(\ref{eq:Af}) that for any (divergence-free) test vector field $f^a$, we have
\begin{equation}
\label{K2pt}
\bra{\Omega} \left[\op{A}_{a}^{\textrm{in}}(f^{a}) \right]^2 \ket{\Omega} = || K \Delta (f)||^2
\end{equation}
where $\op{A}^{\textrm{in}}$ denotes the unperturbed electromagnetic field. Combining Eq.~(\ref{K2pt}) with Eq.~(\ref{eq:entphot}) (with $G^{\textrm{ret}}$ replaced by $-\Delta$), we obtain
\begin{equation}
\label{eq:Nloc}
\braket{N} = \braket{\Omega|\left[\op{A}_{a}^{\textrm{in}}(j^{a}_{1}-j^{a}_{2})\right]^{2}|\Omega}.
\end{equation}
Thus, we see that the prescription for computing the decoherence of Alice's superposition outlined in the bullet points given in the previous section can be equivalently reformulated as follows:
\begin{itemize}

\item We compute the expected currents $j_1^a$ and $j_2^a$ of the components of Alice's superposition. 

\item We compute the two-point function
$\bra{\Omega}\op{A}_a^{\textrm{in}}(x)\op{A}_{a'}^{\textrm{in}}(x')|\ket{\Omega}$ of the unperturbed field in the vacuum state $\ket{\Omega}$.

\item We smear this two-point function in both variables with the test vector field $f^a = j_1^a - j_2^a$. This yields the expected number of entangling photons, $\braket{N}$, and thereby the decoherence, Eq.~(\ref{eq:decAlice2}).

\end{itemize}

The remarkable feature of this reformulation is that it requires only knowledge of the expected currents and the unperturbed two-point function of the quantum field in Alice's lab, i.e., unlike the previous prescription, we do not need to calculate anything about the particle content of the perturbed field at late times. In particular, this explicitly demonstrates that the decoherence can be viewed as a purely local phenomenon occurring entirely in Alice's lab. 

The corresponding result in the linearized gravitational case is 
\begin{equation}
\braket{N} = \braket{\Omega|\left[\op{h}_{ab}^{\textrm{in}}(T^{ab}_{1}-T^{ab}_{2})\right]^{2}|\Omega} 
\label{eq:Ngrav}
\end{equation}
where $T^{ab}_{1} - T^{ab}_{2}$ is the difference in the stress-energy of the components of Alice's particle (also taking into account the tiny correlated motion of Alice's lab that keeps the center of mass fixed). 
Again, the calculation of decoherence is seen to require only a
knowledge of the expected stress-energy of the components of Alice's particle as well as the unperturbed two-point function of the quantum field in Alice's lab, so the decoherence can be viewed as a purely local phenomenon occurring entirely in Alice's lab. 

Note that Eqs.~(\ref{eq:Nloc}) and (\ref{eq:Ngrav}) show that the quantity $\braket{N}$---and hence the corresponding decoherence, $\mathscr D$, given by Eq.~(\ref{eq:decAlice2})---are determined by the \textit{vacuum fluctuations} of the quantum field smeared into the difference of the sources in Alice's lab.

In the next section, we recompute the black hole decoherence Eq.~(\ref{eq:NEMBH}) using our local reformulation. This will enable us to gain further insights into the nature of the decoherence in the presence of a black hole and to compare it with cases where no horizon is present.

\section{Local Calculation of the Decoherence in the Unruh Vacuum Around a Schwarzschild Black Hole}
\label{locsch}

We now compute the decoherence of Alice's particle in the presence of a Schwarzschild black hole by the methods of the previous section. We will focus upon the electromagnetic case and merely comment briefly on the linearized gravitational case near the end of this section.

If we neglect the spatial extent of the particle components, then we have
\be
\label{jpp}
j_1^a (t,x^i) \approx \frac{q}{\sqrt{-g}} \delta^{(3)}[x^i - X_1^i(t)] u_1^a  \frac{d \tau_1}{dt}
\ee
and similarly for $j_2^a$. Here $t$ is the Killing time coordinate, $x^i$ are spatial coordinates on the hypersurfaces $\Sigma_t$ orthogonal to the timelike Killing field $t^a$, $X_1^i (t)$ is the path taken by the center of mass of the first component of the particle, $u_1^a$ is the 4-velocity of that path, $\tau_1$ is the proper time along the path, and $\delta^{(3)}$ is the ``coordinate delta function'' defined so that $\int \delta^{(3)}[x^i - X_1^i(t)] d^3 x = 1$. For nonrelativistic motion relative to the rest frame of $t^a$, we have $d \tau_1/ dt \approx \sqrt{-g_{tt}}$ and 
\be
j_1^a (t,x^i) \approx \frac{q}{\sqrt{-g}} \delta^{(3)}[x^i - X_1^i(t)] (t^a + v_1^a)
\ee
where $v^a$ is the coordinate velocity of the component, i.e., $v_1^i = dX_1^i/dt$ and $v_1^t =0$. 
We represent the displacement of the two components of Alice's particle at time $t$ by the tangent vector $S^a(t)$ to the geodesic segment in $\Sigma_t$ of unit affine parameter that connects the centers of mass of the two components. We write $S^a(t) = d(t) s^a(t)$, where $s^a$ is a unit vector. Then $d(t)$ represents the proper distance between the components. We assume that $s^a$ is Lie transported along $t^a$ (i.e., the direction of separation does not change with time) and that $d(t)$ is smoothly varying and is such that
\begin{equation}
\label{eq:d}
d(t) = \begin{cases}
d \textrm{ for }|t|< T/2 \\
0 \textrm{ for }t<-T/2- T_{1} \textrm{ and } t> T/2 + T_{2}.
\end{cases}
\end{equation}

The difference between the current densities of the two components is given by
\begin{eqnarray}
 (j_1^a - j_2^a) &\approx& \frac{q d(t)}{\sqrt{-g}} t^a s^b \nabla_b  \delta^{(3)}(x^i - X^i) \nonumber \\
&-& \frac{q}{\sqrt{-g}} \delta^{(3)}(x^i - X^i) s^a t^b \nabla_b d(t) 
\end{eqnarray}
where $X^i$ is the position of Alice's lab. Here, the first term arises from the difference in charge densities and the second term arises from the difference in spatial currents. We may rewrite this as
\be
(j_1^a - j_2^a) \approx \frac{2 q}{\sqrt{-g}}  t^{[a} s^{b]} \nabla_b \left[ d(t) \delta^{(3)}(x^i - X^i) \right].
\label{dipsource}
\ee

We define the electric field $E_a$ on the static slices by\footnote{Note that this differs from the notion of the ``electric field on the horizon'' used in \cite{Danielson:2022tdw,Danielson:2022sga}, which was defined as $F_{ab} k^b$, where $k^b$ is the null normal to the horizon.} 
\be
E_a = F_{ab} t^b = (\nabla_a A_b - \nabla_b A_a) t^b \, .
\ee
It follows immediately from Eq.~(\ref{dipsource}) and the definition of $E$ that the unperturbed field $\op{A}^{\textrm{in}}$ smeared in with $j_1^a - j_2^a$ (with the volume element $\sqrt{-g} d^4 x$ understood in the smearing) is given by
\be
 \op{A}_{a}^{\textrm{in}} (j_1^a - j_2^a) \approx - q \int dt d(t) s^a \op{E}_a^{\textrm{in}} (t, X^i). 
\label{Asmear}
\ee
Thus, from Eq.~(\ref{eq:Nloc}), we have
\be
\braket{N} = q^2 \int dt dt' d(t) d(t') \braket{s^a \op{E}_a^{\textrm{in}} (t, X^i) s^{a'} \op{E}_{a'}^{\textrm{in}} (t', X^i)}_{\Omega}.
\label{N2pt}
\ee
Thus, to calculate $\braket{N}$ and thereby the decoherence Eq.~(\ref{eq:decAlice2}) of Alice's particle, we simply evaluate the two-point function of the component, $s^a \op{E}_a^{\textrm{in}}$ of the electric field in the direction of the separation, $s^a$, of the components of Alice's particle evaluated at Alice's lab, $x^i = X^i$, and smeared in time via the separation $d(t)$.

Thus, the remaining task is to obtain the two-point function of the unperturbed electric field, which we will do via a mode expansion. We shall simplify this task by restricting consideration to the case of radial separation of the components of Alice's particle, so that we need only calculate the two-point function of the radial component of $\op{E}_a^{\textrm{in}}$. The magnetic parity modes do not contribute to the radial component of the electric field so we need only consider the electric parity modes \cite{Wald:2022}. The two-point function of the radial coordinate component $\op{E}_r^{\textrm{in}}$ has been calculated for the Boulware, Unruh and Hartle-Hawking vacuum states by Zhou and Yu \cite{Zhou:2012eb} and Menezes \cite{Menezes_2015}, who obtained\footnote{These results are given in Eqs.~(51)-(53) of \cite{Zhou:2012eb} and Eqs.~(A13)-(A16) of \cite{Menezes_2015}. We have used the addition theorem for spherical harmonics to rewrite their sum of spherical harmonics over azimuthal number $m$ in terms of $P_{\ell}(\hat{r}\cdot \hat{r}^{\prime})$.}
\begin{align}
\label{eq:Er2pt}
&\braket{\op{E}_{r}(x)\op{E}_{r}(x^{\prime})}_{\Omega} =  \sum_{\ell=1}^{\infty}  \frac{C_{\ell}P_{\ell}(\hat{r}\cdot \hat{r}^{\prime})}{\textbf{}r^{2}r^{\prime 2}}\int_{-\infty}^{\infty} \frac{ d\omega}{\omega }e^{-i\omega (t-t^{\prime})}\times \nonumber \\
&\times \bigg[\vec{G}(\omega)\vec{R}_{\omega \ell}(r)\vec{R}^*_{\omega \ell}(r^{\prime})+ \cev{G}(\omega)\cev{R}_{\omega \ell}(r) \cev{R}^*_{\omega \ell}(r^{\prime})\bigg].
\end{align}
Here,
\begin{equation}
C_{\ell} \equiv \frac{1}{16\pi^2}\ell(\ell +1)(2\ell +1)
\end{equation}
and $P_\ell$ is the $\ell$th Legendre polynomial (so $P_{\ell}(\hat{r}\cdot \hat{r}^{\prime}) = 1$ for the case of interest below where  $x^i= {x'}^i$). The mode functions $\vec{R}_{\omega \ell}(r)$ and $\cev{R}_{\omega \ell}(r)$ satisfy the differential equation 
\begin{equation}
\label{eq:Rdiff}
\frac{d^{2}R_{\omega\ell }}{dr^{\ast 2}} + \bigg[\omega^2-V(r)\bigg]R_{\omega\ell } = 0
\end{equation}
where 
\begin{equation}
\label{eq:potential}
V(r)=\left(1-\frac{2M}{r}\right) \frac{\ell(\ell+1)}{r^2}
\end{equation}
and $r^{\ast}$ is the radial ``tortoise coordinate'' 
\begin{equation}
r^* = r + 2M \ln\left(\frac{r}{2M} - 1 \right) \, ,
\end{equation}
which satisfies $dr^{\ast}/dr = (1-2M/r)^{-1}$ and ranges from $r^* \to -\infty$ at the horizon to $r^* \to +\infty$ at infinity. The modes $\vec{R}_{\omega \ell}$ correspond to waves that are incoming from the white hole and are defined by the asymptotic conditions
\begin{equation}
\label{eq:WHbcs}
\vec{R}_{\omega \ell}(r) \to 
\begin{cases}
e^{i\omega r^{\ast}} + \vec{A}_{\omega \ell}e^{-i\omega r^{\ast}} \textrm{ as $r\to 2M$}\\
\vec{B}_{\omega \ell }e^{i\omega r^{\ast}} \textrm{ as }r\to \infty
\end{cases}
\end{equation}
whereas the modes $\cev{R}_{\omega \ell}$ correspond to waves that are incoming from infinity and are defined by the asymptotic conditions
\begin{equation}
\label{eq:scribcs}
\cev{R}_{\omega \ell}(r) \to 
\begin{cases}
\cev{B}_{\omega \ell}e^{-i\omega r^{\ast}}  \textrm{ as $r\to 2M$}\\
e^{-i\omega r^{\ast}}+ \cev{A}_{\omega \ell }e^{i\omega r^{\ast}} \textrm{ as }r\to \infty.
\end{cases}
\end{equation}
Finally, the coefficients $\vec{G}(\omega)$ and $\cev{G}(\omega)$ appearing in Eq.~(\ref{eq:Er2pt}) depend on the choice of vacuum state $\ket{\Omega}$. For the Boulware vacuum \cite{boulware1975quantum}, $\ket{\Omega_{\rm B}}$, we have
\begin{equation}
\label{eq:Boul}
\vec{G}_{\rm B}(\omega) = {\cev{G}}_{\rm B}(\omega) = \Theta(\omega)
\end{equation}
corresponding to the fact that Boulware vacuum is positive frequency with respect to Killing time at both the white hole horizon and past infinity. For the Unruh vacuum \cite{PhysRevD.14.870}, $\ket{\Omega_{\rm U}}$, we have
\begin{equation}
\label{eq:Unruh}
\vec{G}_{\rm U}(\omega) = \frac{1}{1-e^{-2\pi \omega/\kappa}} \quad \textrm{ and } \quad {\cev{G}}_{\rm U}(\omega) = \Theta(\omega) 
\end{equation}
where $\kappa$ is the surface gravity of the black hole,
corresponding to the fact that the Unruh vacuum is positive frequency with respect to Killing time at past null infinity but is positive frequency with respect to affine time (and thus is thermally populated with respect to Killing time at temperature $\kappa/2 \pi$) on the white hole horizon. 
Finally, for the Hartle-Hawking vacuum \cite{Hartle:1976tp}, $\ket{\Omega_{\rm HH}}$, we have
\begin{equation}
\label{eq:HH}
\vec{G}_{\rm HH}(\omega) = {\cev{G}}_{\rm HH}(\omega)=\frac{1}{1-e^{-2\pi \omega/\kappa}} 
\end{equation}
corresponding to the fact that the Hartle-Hawking vacuum is a thermal state at both the white hole horizon and past null infinity.

We now plug our expression Eq.~(\ref{eq:Er2pt}) for the two-point function into our formula Eq.~(\ref{N2pt}) for $\braket{N}$. We obtain
\begin{align}
\label{eq:Er2ptN}
\braket{N} &= q^2 \sum_{\ell=1}^{\infty}  \frac{C_{\ell}(1-2M/r)}{r^4} \int_{-\infty}^{\infty} \frac{ d\omega}{\omega } |\hat{d}(\omega)|^2\times \nonumber \\
&\times \bigg[\vec{G}(\omega)|\vec{R}_{\omega \ell}(r)|^{2}
+ \cev{G}(\omega)|\cev{R}_{\omega \ell}(r)|^{2}
\bigg].
\end{align}
Here $r$ is the radial coordinate of Alice's lab and $\hat{d}(\omega)$ is  the Fourier transform of $d(t)$
\begin{equation}
\hat{d}(\omega) = \int_{-\infty}^{\infty}dt~e^{i\omega t}d(t) \, .
\end{equation}
The factor of $(1-2M/r)$ arises from converting the proper distance component $s^a E_a$ appearing in Eq.~(\ref{N2pt}) to the coordinate component $E_r$ appearing in Eq.~(\ref{eq:Er2pt}), and we used the fact that $P_{\ell}(1)=1$. 

For $d(t)$ of the form Eq.~(\ref{eq:d}) with $T$ large, the magnitude of the Fourier transform $|\hat{d}(\omega)|$ behaves like $d/|\omega|$ as $\omega \to 0$ until this divergent behavior levels off below $|\omega| \sim 1/T$. There will also be a high frequency cutoff at $|\omega| \sim 1/{\rm min} [T_1, T_2]$. Thus, we may approximate the contribution of $|\hat{d}(\omega)|$ to the integral in Eq.~(\ref{eq:Er2ptN}) using
\begin{equation}
\label{dFT}
|\hat{d}(\omega)| \sim 
\begin{cases}
\frac{d}{\omega} \quad \quad  \frac{1}{T} < |\omega| < \frac{1}{{\rm min} [T_1, T_2]} \\
0 \quad \quad |\omega| <  \frac{1}{T} \,\,\, \textrm{or} \,\,\, |\omega| > \frac{1}{{\rm min} [T_1, T_2]}.
\end{cases}
\end{equation}
Thus, the behavior of $\braket{N}$ at large $T$ will be determined by the behavior of the integrand of Eq.~(\ref{eq:Er2ptN}) near the low frequency end, $|\omega| \sim 1/T$, of the range of integration. In order to determine this behavior, we need to obtain expressions for the mode functions $\vec{R}_{\omega \ell}(r)$ and $\cev{R}_{\omega \ell}(r)$ at very low frequencies.

\begin{figure}
\label{fig:potential}
\includegraphics[width=\columnwidth]{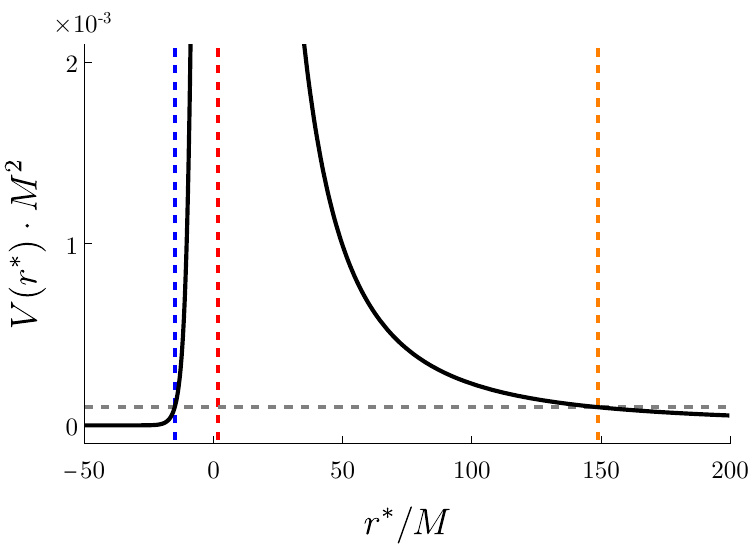}
    \caption{The potential $V(r^{\ast})$ plotted as a function of $r^{\ast}$ for $\ell=1$. The horizontal, grey dashed line corresponds to square of the frequency $\omega= 0.01/M$. The vertical blue and orange dashed lines correspond to the turning points $r^{\ast}_{1}$ and $r^{\ast}_{2}$ respectively. The vertical, red dashed line is the peak of the potential at $r=3M$. The radial mode solutions in regions II and III are matched in the regions where they overlap. The solutions in regions I and II are both good approximations in a neighborhood of $r^{\ast}=r^{\ast}_{1}$ and so can be matched there.}
\end{figure}

In order to determine these mode functions at low frequencies, we divide the exterior into three regions (see Fig.~\ref{fig:potential}):
\begin{eqnarray}
\textrm{Region} &\textrm{I}& \quad 2M < r \leq r_1 \\
\textrm{Region} &\textrm{II}& \quad r_1 < r \ \ll\ r_2 \\
\textrm{Region} &\textrm{III}& \quad 3M \ll r < \infty
\end{eqnarray}
where \cite{Fabbri_1975}
\be
r_1 = 2M + \frac{8\omega^2 M^3}{\ell(\ell +1)}
\ee
\be
r_2 = \frac{[\ell(\ell +1)]^{1/2}}{\omega}.
\ee
Note that for $\omega M \ll 1$, there will be a large overlap of regions II and III. In region I, we may neglect the potential, $V(r)$, in Eq.~(\ref{eq:potential}) compared with $\omega^2$ and the solutions take the form
\be
R_{\omega \ell}^{\textrm{I}}(r)  \approx \alpha^\textrm{I}_\ell(\omega )  e^{i \omega r^*} + \beta^\textrm{I}_\ell(\omega )  e^{- i \omega r^*} \, .
\label{R1}
\ee
In region II, the potential, $V(r)$, dominates over $\omega^2$ and the solutions are well approximated by the static (zero frequency) solutions \cite{Cohen_1971,Fabbri_1975}
\begin{align}
\label{eq:zerofreq}
R_{\omega \ell}^{\textrm{II}}(r)  \approx& \alpha^{\textrm{II}}_{\ell}( \omega ) \bigg[\frac{y}{2}P_{\ell}(y-1) - \frac{P_{\ell+1}(y-1)-P_{\ell-1}(y-1)}{2(2\ell+1)}\bigg]\nonumber \\
+\beta^{\textrm{II}}_{\ell} (\omega )& \bigg[\frac{y}{2}Q_{\ell}(y-1) - \frac{Q_{\ell+1}(y-1)-Q_{\ell-1}(y-1)}{2(2\ell+1)}\bigg]
\end{align}
where $y \equiv r/M$. 

Finally, in region III, we may approximate the potential as $V(r) \approx \ell(\ell +1)/r^{*2}$ and we may then approximate the solutions by the flat spacetime solutions with $r^*$ replacing $r$
\be
R_{\omega \ell}^{\textrm{III}}(r)  \approx \alpha{^\textrm{III}_\ell}(\omega )  r^{\ast}j_{\ell}(\omega r^*) + \beta{^{\textrm{III}}_\ell} (\omega )  r^* n_{\ell}(\omega r^*) 
\label{R3}
\ee
where $j_\ell$ and $n_\ell$ denote the spherical Bessel and Neumann functions. Note that in the overlap between regions II and III, we may neglect\footnote{Replacement of $r^*$ by $r$ in Eq.~(\ref{R3}) would give rise to an arbitrarily large phase error in the solutions as $r \to \infty$, so the difference between $r$ and $r^*$ cannot be neglected throughout region III. However, the difference between $r$ and $r^*$ makes only a small correction, which we neglect, in the overlap of regions II and III.} the difference between $r$ and $r^*$ and the solutions take the form 

\be
R_{\omega \ell}^{\textrm{II,III}}(r)  \approx \alpha_{\ell}( \omega) r^{\ell + 1} + \frac{\beta_{\ell}(\omega)}{r^{\ell}}.
\ee

In order to determine $\vec{R}_{\omega \ell}(r)$, we start with the solution $\vec{B}_{\omega \ell}e^{-i\omega r^{\ast}}$ in region III [see Eq.~(\ref{eq:WHbcs})], with initially unknown coefficient $\vec{B}_{\omega \ell}$. We match this solution to the general solution Eq.~(\ref{eq:zerofreq}) in region II and then match the resulting solution to the general solution Eq.~(\ref{R1}) in region I. Finally, we adjust $\vec{B}_{\omega \ell}$ so as to give a coefficient of $1$ to the term $e^{i \omega r{^*}}$ as $r \to 2M$ in Eq.~(\ref{eq:WHbcs}). Similarly, to obtain $\cev{R}_{\omega \ell}(r)$, we start with the solution $\cev{B}_{\omega \ell}e^{-i\omega r^{\ast}}$ in region I 
[see Eq.~(\ref{eq:scribcs})], with initially unknown coefficient $\cev{B}_{\omega \ell}$. We match this solution to the general solution Eq.~(\ref{eq:zerofreq}) in region II, match the resulting solution to the general solution Eq.~(\ref{R3}) in region III, and adjust $\cev{B}_{\omega \ell}$ so as to give a coefficient of $1$ to the term $e^{-i\omega r^{\ast}}$ as $r \to \infty$ in Eq.~(\ref{eq:scribcs}). 

For simplicity, we shall assume that Alice's lab is located in the region $M\ll r\ll 1/\omega$ for the relevant range of frequencies $\omega\sim 1/T$, so that it lies in the overlap of regions II and III. This is the regime in which the estimates of \cite{Danielson:2022tdw,Danielson:2022sga} reviewed in Sec. \ref{sec:gedanken} apply, so we will be able to make a direct comparison of our results with the results of the previous calculation. The mode functions $\cev{R}_{\omega \ell}(r)$ were previously obtained by Fabbri \cite{Fabbri_1975}, since they are needed to analyze scattering of classical waves by a black hole. In region III, we find that $\beta^{\textrm{III}}_\ell = O([\omega M]^{2\ell + 2})$ and thus the Neumann term in Eq.~(\ref{R3}) may be neglected. The solution with the correct normalization in region III is 
\begin{equation}
\cev{R}_{\omega \ell}(r) \approx -2 i^{3 \ell+1} \omega r^* j_{\ell}(\omega r^*).
\label{infmodes}
\end{equation}
If, in addition, we have $\omega r \ll 1$, then 
\begin{equation}
\label{eq:Rinfapprox}
 \cev{R}_{\omega \ell}(r)  \approx -\frac{i^{3\ell +1}2^{\ell+1}\ell!}{(2\ell+1)!}  (\omega r)^{\ell+1} \quad \textrm{ $(M\ll r\ll \omega^{-1})$}.
\end{equation}
Thus, as might be expected, if we assume that Alice's lab is not close to the black hole ($r \gg M$), the modes in Alice's lab corresponding to low frequency incoming waves from infinity are essentially unaffected by the black hole. As in flat spacetime, they are suppressed by the factor $(\omega r)^{\ell + 1}$ due to the angular momentum barrier. Since $\omega r \ll 1$, the dominant contribution to the two-point function in Alice's lab from modes that are incoming from infinity arises from the $\ell = 1$ mode.

Performing the similar analysis for $\vec{R}_{\omega \ell}(r)$, we obtain 
\begin{equation}
\label{eq:RHW}
\vec{R}_{\omega \ell}(r)  \approx a_{\ell}\bigg(\frac{M}{r}\bigg)^{\ell}(M\omega)\quad \textrm{ $(M\ll r\ll \omega^{-1})$},
\end{equation}
where 
\begin{equation}
a_{\ell}={\frac{-i 2^{l+2}   (\ell-1)! (\ell+1)! }{(2 \ell+1) (2 \ell)!}}.
\end{equation}
Note that, although at low frequencies the white hole modes are essentially entirely reflected back into the black hole by the potential barrier $V(r)$, these modes fall off in $r$ only as the power law $1/r^\ell$ and, thus, they penetrate far beyond the peak of the potential barrier at $r = 3M$ and can have a nontrivial effect in Alice's lab. Note also that, as opposed to the incoming modes from infinity, the frequency dependence of the white hole modes is $\ell$ independent. Since $r \gg M$, the dominant contribution to the two-point function in Alice's lab from the modes emerging from the white hole arises from the $\ell = 1$ modes.

We now estimate $\braket{N}^{\rm U}$, Eq.~(\ref{eq:Er2ptN}), for the case of the Unruh vacuum, $\ket{\Omega_{\rm U}}$. (The cases of the Boulware and Hartle-Hawking vacua will be treated in the next section.) We first consider the contribution, $\braket{N}_{\leftarrow}^{\rm U}$, of the incoming modes from infinity. We keep only the $\ell = 1$ contribution and use Eq.~(\ref{eq:Rinfapprox}) to evaluate $\cev{R}_{\omega 1}$. We use Eq.~(\ref{dFT}) to evaluate $\hat{d}$ and we also use ${\cev{G}}_{\rm U}(\omega) = \Theta(\omega)$.  Ignoring all subleading terms and all factors of order unity, we obtain the following expression for the contribution of the incoming modes from infinity in the Unruh vacuum
\begin{equation}
\label{eq:N2}
\braket{N}_{\leftarrow}^{\rm U} \sim \frac{q^{2}}{r^4}\int_{1/T}^{1/\textrm{min[$T_{1},T_{2}$]}} \frac{d\omega}{\omega} \frac{d^2}{\omega^2} (\omega r)^4   \sim \frac{q^{2}d^{2}}{\textrm{min}[T_{1},T_{2}]^{2}} \, .
\end{equation}
This agrees with the estimate Eq.~(\ref{eq:NEMMink}) for Minkowski spacetime obtained by considering radiation of entangling photons to infinity. Note that the contribution from the incoming modes from infinity does not grow with $T$. 

Next, we estimate the contribution, $\braket{N}_{\to}^{\rm U}$ of the incoming modes from the white hole to $\braket{N}^{\rm U}$. We keep only the $\ell = 1$ contribution and use Eq.~(\ref{eq:RHW}). In the Unruh vacuum, we have
\begin{equation}
\label{eq:thermal}
\vec{G}_{\rm U}(\omega) = \frac{1}{1-e^{-2\pi \omega/\kappa}} \approx \frac{\kappa}{2\pi \omega}.
\end{equation}
Ignoring all subleading terms and all factors of order unity and setting $r=D$, we obtain the following expression for the contribution of the incoming modes from the white hole in the Unruh vacuum
\begin{equation}
\braket{N}_{\rightarrow }^{\rm U} \sim \frac{q^{2}d^{2}\kappa M^{4}}{D^{6}} \int_{1/T}^{1/\textrm{min[$T_{1},T_{2}$]}}\frac{d\omega}{\omega^{2}} \sim \frac{q^2 d^2 M^3}{D^6} T.
\label{Nunr}
\end{equation}
For large $T$, this contribution dominates over Eq.~(\ref{eq:N2}), so we have
\be
\braket{N}^{\rm U} = \braket{N}_{\leftarrow }^{\rm U} + \braket{N}_{\rightarrow }^{\rm U} \approx \braket{N}_{\rightarrow }^{\rm U} \sim \frac{q^2 d^2 M^3}{D^6} T.
\label{Nunvac}
\ee
This agrees with the estimate Eq.~(\ref{eq:NEMBH}) for the decoherence resulting from the emission of entangling photons through the black hole horizon. Thus, our purely local analysis reproduces the results previously obtained in~\cite{Danielson:2022tdw,Danielson:2022sga}.

 We now briefly comment on the analogous computation in the linearized quantum gravitational case. If we approximate the stress-energy tensor of the first component of Alice's particle as being essentially a point particle, then its stress-energy tensor would take the form
\be
T_1^{ab} (t) \approx \frac{m}{\sqrt{-g}} \delta^{(3)}[x^i - X_1^i(t)] u_1^a u_1^b \frac{d \tau_1}{dt}
\ee
in analogy with Eq.~(\ref{jpp}). If this component was not interacting with any other matter, then conservation of stress-energy would imply that it must move on a geodesic. However, since we want the component to follow a nongeodesic trajectory, Alice must apply some ``external force'' to it. The external forces on the different components act oppositely on the different components during separation and recombination and will have a backreaction effect on Alice's lab. In Minkowski spacetime, conservation of total stress-energy implies that Alice's lab would have to move oppositely to the particle components so as to keep the center of mass of the total system fixed. In the case of a black hole spacetime, the situation is more complicated, since a further external system would be needed to keep Alice's lab stationary. Nevertheless, the analog of the dipole contribution Eq.~(\ref{dipsource}) to the difference in stress-energy of the components should be canceled by the stress-energy effects of Alice's lab, and the leading order contribution should be given by 
\be
(T_1^{ab} - T_2^{ab}) \approx \frac{2 m}{\sqrt{-g}} \frac{dt}{d\tau} t^{[a} s^{c]} t^{[b} s^{d]} \nabla_c \nabla_d\left[ d^2(t) \delta^{(3)}(x^i - X^i) \right].
\label{quadsource}
\ee
The analog of Eq.~(\ref{Asmear}) is then
\be
 \op{h}_{ab}^{\textrm{in}} (T_1^{ab} - T_2^{ab}) \approx -m \int dt d^2(t) s^a s^b \op{E}_{ab}^{\textrm{in}} (t, X^i). 
\label{hsmear}
\ee
where $\op{E}_{ab}^{\textrm{in}}$ is the quantum field observable corresponding to the electric part of the 
Weyl tensor $E_{ab} = C_{acbd}t^{c}t^{d}$. Thus, the computation of $\braket{N}$, Eq.~(\ref{eq:Ngrav}), reduces to obtaining the two-point function of the Weyl tensor. Again, we can simplify calculations by restricting to the case of radial separation. The upshot is that the order of magnitude estimates that we obtained above for the electromagnetic case apply with the substitutions $q\to m$, ${d} \to {d^{2}}$ and the mode sum now running over $\ell\geq 2$, so that the dominant contribution arises from $\ell = 2$. For the Unruh vacuum, this yields the estimate 
\be
\braket{N}_{\leftarrow}^{\rm U, GR} \sim \frac{m^{2}d^{4}}{\textrm{min}[T_{1},T_{2}]^{4}}
\label{locUbh1}
\ee
in agreement with Eq.~(\ref{eq:NGRMink}), and the estimate
\be
\braket{N}_{\rightarrow }^{\rm U, GR} \sim \frac{M^{5}m^{2}d^{4}}{D^{10}}T 
\label{locUbh2}
\ee
in agreement with Eq.~(\ref{eq:NGRBH}). 

Finally, we note that Eq.~(\ref{N2pt}) shows that in the electromagnetic case, we have 
\be
\braket{N} = q^2 \left\langle\left(\int dt d(t) s^a\op{E}_a^{\rm in} \right)^2\right\rangle_\Omega \sim q^2 d^2 T^{2} \left[\Delta (s^a \op{E}_a^{\rm in})\right]^2
\ee
where $\Delta (s^a \op{E}_a^{\rm in})$ is defined by 
\be
[\Delta (s^a \op{E}_a^{\rm in})]^2 = \left\langle\left(\frac{1}{T}\int dt \frac{d(t)}{d} s^a\op{E}_a^{\rm in} \right)^2\right\rangle_\Omega
\label{Efluct}
\ee
and thus can be interpreted as the root mean square of the time average of the $s^a$ component of the electric field fluctuations in state $\ket{\Omega}$ on Alice's worldline during the duration of her experiment. 

The fluctuations of the electric field are most usefully characterized by its power spectrum. The power spectrum of the radial component of the electric  $S^{\textrm{U}}_{r}(\omega)$ is given by
\be
S^{\textrm{U}}_{r}(\omega) = \int_{-\infty}^{\infty}dt~e^{i\omega (t-t^{\prime})}\braket{\op{E}_{r}(t,X^{i})\op{E}_{r}(t^{\prime},X^{i})}_{\Omega_{\textrm{U}}}.
\ee
The modes that dominantly contribute to this power spectrum in Alice's lab are the white hole modes $\vec{R}_{\omega \ell}$ with $\ell =1$ and $\omega \sim 1/T$. By Eqs.~(\ref{eq:Er2pt}) and (\ref{eq:RHW}), in the Unruh vacuum these modes contribute\footnote{In Rindler spacetime, the analogous horizon modes similarly make a contribution to the power spectrum of the electric field that is nonvanishing as $\omega \to 0$ \cite{Wilson-Gerow:2024ljx}. This fact is undoubtedly intimately related to the phenomena analyzed in \cite{Higuchi:1992we,Higuchi:1992td,Matsas:1996yh,Higuchi:1996aj}.} 
\begin{align}
S^{\textrm{U}}_{r}(\omega) 
\sim & ~\frac{1}{r^4} \frac{1}{\omega} \vec{G}_{U}(\omega) |\vec{R}_{\omega 1}(r)|^2 \nonumber \\
\sim&~\frac{\kappa}{r^4 \omega^2} \left(\frac{M^2 \omega}{r}\right)^2 \nonumber \\
\sim &~\frac{M^{3}}{r^{6}} .
\end{align}
This corresponds to the black hole in the Unruh vacuum acting as though it were an ordinary body with a randomly fluctuating electric dipole moment, $\vec{P}_{U}$ with constant power spectrum
\begin{equation}
\label{fluctdip}
\Delta |\vec{P}_{\textrm{U}}|(\omega)\sim \frac{\sqrt{\epsilon_{0}\hbar}G^{3/2}M^{3/2}}{c^{3}} \sim 10~\frac{\textrm{e$\cdot$m}}{\sqrt{\textrm{Hz}}}~\bigg(\frac{M}{M_{\odot}}\bigg)^{3/2},
\end{equation}
where we have restored fundamental constants to emphasize that this is an $O(\sqrt{\hbar})$ effect.

Similarly, in the gravitational case, the black hole acts as though it were an ordinary body with a fluctuating mass quadrupole moment of magnitude
\be
\Delta |Q_{\textrm{U}}|(\omega) \sim \frac{\sqrt{\hbar} G^{2}M^{5/2}}{c^{5}}\sim 10^{-1} ~\frac{\textrm{g$\cdot$m$^{2}$}}{\sqrt{\textrm{Hz}}}~\bigg(\frac{M}{M_{\odot}}\bigg)^{5/2} \, .
\label{fluctquad}
\ee

More generally, the power spectra of the higher electric multipole fluctuations and mass multipole fluctuations of the black hole go as 
\be 
\Delta|\ms{Q}_{\ell}^{\EM}|(\omega) \sim M^{\ell + 1/2}, \quad \Delta|\ms{Q}_{\ell}^{\GR}|(\omega) \sim M^{\ell + 1/2}.
\ee

There also are similar fluctuations of the magnetic parity multipole moments. The dominant contribution to the decoherence in Alice's experiment, however, comes from the lowest electric parity multipole moment.

In conclusion, we have successfully reproduced the main results of \cite{Danielson:2022tdw,Danielson:2022sga} using our purely local reformulation. In the next section, we will use our local reformulation to compare the results for the decoherence in the Unruh vacuum around a black hole to other cases.

\section{Comparison with decoherence arising in other cases}
\label{sec:comp}
The results we have obtained in the previous section will now enable us to analyze the decoherence arising in other situations. Specifically, we will analyze the cases of (i) a Schwarzschild black hole in the Boulware or Hartle-Hawking vacuum, (ii) Minkowski spacetime in the Minkowski vacuum or filled with a thermal bath of radiation, (iii) a spacetime corresponding to the gravitational field of a star with no internal degrees of freedom assigned to the star, and (iv) a material body with internal degrees of freedom in a thermal state.

\subsection{Decoherence in the Boulware and Hartle-Hawking Vacua}

The Boulware vacuum, $\ket{\Omega_{\rm B}}$, is the ground state for the exterior region ($r > 2M$) of Schwarzschild with respect to the timelike Killing field. The Boulware vacuum is singular on the past and future event horizons of Schwarzschild. Since it is singular on the future horizon, it does not correspond to a physically reasonable state for a black hole formed by gravitational collapse. Nevertheless, the Boulware vacuum is a well-defined state in Alice's lab, and it is instructive to compute the decoherence of her particle in the Boulware vacuum using the results of the previous section.

The Boulware vacuum differs from the Unruh vacuum only in that $\vec{G}$ and $\cev{G}$ are now given by Eq.~(\ref{eq:Boul}) rather than Eq.~(\ref{eq:Unruh}). Since $\cev{G}_{\rm B} = \cev{G}_{\rm U}$, it follows immediately that $\braket{N}_{\leftarrow}^{\rm B}$ is again given by Eq.~(\ref{eq:N2}), i.e.,
\begin{equation}
\label{eq:N4}
\braket{N}_{\leftarrow}^{\rm B} = \braket{N}_{\leftarrow}^{\rm U}  \sim \frac{q^{2}d^{2}}{\textrm{min}[T_{1},T_{2}]^{2}} \, .
\end{equation}
On the other hand, in the Boulware vacuum, we have $\vec{G}_{\rm B} = \Theta(\omega)$ rather than being given by Eq.~(\ref{eq:thermal}). Consequently, the integrand of the formula for $\braket{N}_{\rightarrow}^{\rm B}$ will differ from the integrand appearing on the right side of Eq.~(\ref{Nunr}) by a factor of $\sim \omega/\kappa$. We obtain
\begin{eqnarray}
\label{eq:Nright1}
\braket{N}_{\rightarrow}^{\rm B} &\sim& \frac{q^{2}d^{2}M^{4}}{D^{6}}\int_{1/T}^{1/\textrm{min[$T_{1},T_{2}$]}}\frac{d\omega}{\omega} \nonumber \\
&=& \frac{q^{2}d^{2}M^{4}}{D^{6}} \ln\bigg(\frac{T}{\textrm{min[$T_{1},T_{2}$]}}\bigg).
\end{eqnarray}
Additionally, we note that the Boulware vacuum at $M\omega \ll 1$ has a randomly fluctuating electric dipole $\Delta |\vec{P}_{\textrm{B}}|$ and mass quadrupole $\Delta |Q_{\textrm{B}}|$ of magnitude
\begin{equation}
\Delta |\vec{P}_{\textrm{B}}|(\omega) \sim M^{2} \sqrt{\omega} , \quad \Delta |Q_{\textrm{B}}|(\omega)\sim M^{3}\sqrt{\omega}
\end{equation}
which are much smaller than the corresponding fluctuations in the Unruh vacuum given by Eqs.~(\ref{fluctdip}) and (\ref{fluctquad}).

Equation~(\ref{eq:Nright1}) could also be derived by the methods used in \cite{Danielson:2022tdw,Danielson:2022sga}. Indeed, the only change that needs to be made to the calculations done in \cite{Danielson:2022tdw,Danielson:2022sga} is that when we compute the one-particle norm corresponding to the retarded solution with source $j_1^a - j_2^a$ on the horizon, we now have to take the positive frequency part with respect to Killing time rather than affine time. The same calculation as led to Eq.~(13) of \cite{Danielson:2022tdw}---which yielded $\braket{N}$ varying as $\ln V$, where $V$ denotes the affine time duration of the separation---now yields the $\ln T$ dependence\footnote{Affine time $V$ is related to Killing time $T$ by $V \propto \exp(\kappa T)$, so, for the Unruh vacuum, the logarithmic dependence on $V$ is converted to the linear dependence on $T$ obtained above. However, for an extremal black hole ($\kappa = 0$), the relation between $V$ and $T$ is linear, so one would expect only logarithmic growth of $\braket{N}$ with $T$ in the extremal case. In fact, in the electromagnetic case, the coefficient of this logarithmic term also vanishes in extremal Kerr \cite{Gralla:2023oya} (the ``black hole Meisner effect'') but a $\ln T$ dependence occurs for a scalar field \cite{Gralla:2023oya}.} given in Eq.~(\ref{eq:Nright1}). 

Next, we consider decoherence in the Hartle-Hawking vacuum, $\ket{\Omega_{\rm HH}}$. In the exterior region ($r > 2 M$) of Schwarzschild, the Hartle-Hawking vacuum is a thermal (KMS) state with respect to all modes at temperature ${\mathscr T} = \kappa/2 \pi$.
Since $\vec{G}_{\rm HH} = \vec{G}_{\rm U}$, it follows immediately that $\braket{N}_{\rightarrow}^{\rm HH}$ is again given by Eq.~(\ref{Nunr}), i.e.,
\begin{equation}
\braket{N}_{\rightarrow }^{\rm HH} = \braket{N}_{\rightarrow }^{\rm U} \sim \frac{q^2 d^2 M^3}{D^6} T.
\label{NHH}
\end{equation}
On the other hand, in the Hartle-Hawking vacuum we have 
\begin{equation}
\label{eq:thermal2}
\cev{G}_{\rm HH}(\omega) = \frac{1}{1-e^{-\omega/{\mathscr T}}}
\end{equation}
with ${\mathscr T} = \kappa/2 \pi = 1/8\pi M$ rather than $\cev{G} = \Theta(\omega)$ as for the Boulware and Unruh vacua. At low frequencies, we have $\cev{G}_{\rm HH}(\omega) \approx {\mathscr T}/\omega$.
Consequently, the integrand (of the formula for $\braket{N}_{\leftarrow}^{\rm HH}$ will differ from the integrand appearing on the right side of Eq.~(\ref{eq:N2}) by a factor of ${\mathscr T}/\omega$ at low frequencies. We obtain
\begin{equation}
\label{eq:NHH}
\braket{N}_{\leftarrow}^{\rm HH} \sim \frac{q^{2}d^{2} {\mathscr  T}}{\textrm{min}[T_{1},T_{2}]} \sim \frac{q^{2}d^{2}}{M \textrm{min}[T_{1},T_{2}]} \, ,
\end{equation}
which differs from Eq.~(\ref{eq:N2}) in that a factor of $M$ has replaced a factor of $\textrm{min}[T_{1},T_{2}]$ in the denominator. Nevertheless,
the thermal population of incoming modes from infinity does not lead to a decoherence that grows with $T$. The key point is that although the radiation incoming from infinity is thermal, it does not have the necessary population of ``soft modes'' to provide a decoherence effect similar to the white hole modes \cite{Wilson-Gerow:2024ljx}.
For sufficiently large $T$ the contribution of the incoming modes from infinity will be negligible compared with the contribution from the white hole modes, Eq.~(\ref{NHH}), and the decoherence in the Hartle-Hawking vacuum will be the same as in the Unruh vacuum.  

It should be noted that there can be additional decoherence effects resulting from thermal populations of modes emerging from the white hole and/or infinity that have not been taken into account in our analysis above. In particular, we have implicitly assumed in our analysis that the components of Alice's particle move on fixed trajectories that are not affected by the incoming radiation. This would be the case if, e.g., the components of Alice's particle are rigidly held in traps.\footnote{It would be best to use nonelectromagnetic traps, so that the traps do not produce any shielding or other electromagnetic effects that could interfere with Alice's experiment.} However, if these components are free to move in response to the incoming electromagnetic radiation, there will be Thompson scattering of the radiation. Since the Thompson scattering will be slightly different for the different components, this will result in decoherence that will grow with time for a steady influx of radiation. The decoherence arising from Thompson scattering of low frequency thermal radiation was estimated in \cite{Danielson:2022sga}, based upon previous analyses of collisional decoherence given in \cite{collisional_Diosi,collisional_Gallis,collisional_Hornberger,collisional_Joos}. It was shown in \cite{Danielson:2022sga} that in the Rindler case, this collisional decoherence can be neglected compared with the decoherence due to emission of soft radiation. For the case of a black hole in the Unruh or Hartle-Hawking states, the same would be true if Alice's lab is sufficiently near the black hole. However, the decoherence rate due to emission of soft radiation falls off rapidly with distance, $D$, from the black hole, whereas the collisional decoherence rate falls off more slowly in the Unruh vacuum and does not fall off at all in the Hartle-Hawking vacuum. Thus, if the particle components are free to respond to the incoming radiation, the collisional decoherence effects will dominate at sufficiently large distances from the black hole.

Finally, we briefly mention the corresponding results for the gravitational case.
In the gravitational case, a calculation analogous to that which led to Eq.~(\ref{eq:Nright1}) now yields 
\be
\label{eq:Nright2}
\braket{N}_{\rightarrow}^{\rm B, GR} \sim \frac{m^{2}d^{4}M^{6}}{D^{10}} \ln\bigg(\frac{T}{\textrm{min[$T_{1},T_{2}$]}}\bigg)
\ee
whereas $\braket{N}_{\leftarrow }^{\rm B, GR}$ is the same as for the Unruh vacuum, Eq.~(\ref{locUbh1}).
A calculation analogous to that which led to Eq.~(\ref{eq:NHH}) now yields
\begin{equation}
\label{eq:NHH2}
\braket{N}_{\leftarrow}^{\rm HH, GR}  \sim\frac{m^{2}d^{4} {\mathscr  T}}{\textrm{min}[T_{1},T_{2}]^{3}} \sim \frac{m^{2}d^{4}}{M \textrm{min}[T_{1},T_{2}]^{3}}
\end{equation}
whereas $\braket{N}_{\rightarrow}^{\rm HH, GR}$ is the same as for the Unruh vacuum, Eq.~(\ref{locUbh2}).

\subsection{Decoherence in Minkowski Spacetime}
\label{minkowdec}

In Minkowski spacetime, there are no ``white hole modes,'' $\vec{R}_{\omega \ell}(r)$, of the quantum field. The incoming modes from infinity, $\cev{R}_{\omega \ell}(r)$, are given by
\be
\cev{R}_{\omega \ell}(r) = -2 i^{3 l+1}\omega r j_\ell (\omega r) \, ,
\label{inmink}
\ee
corresponding to taking the limit as $M \to 0$ of the Schwarzschild modes. The
two point function of the radial component of the electric field can be obtained from Eq.~(\ref{eq:Er2pt}) by deleting the white hole modes and using Eq.~(\ref{inmink}) for the incoming modes from infinity.
The Minkowski vacuum, $\ket{\Omega_M}$, corresponds to $\cev{G}(\omega) = \Theta(\omega)$. It follows immediately that the decoherence of Alice's particle in the Minkowski vacuum will be given by the same estimate as we previously obtained for the decoherence effects of the incoming modes from infinity in Schwarzschild for the Boulware or Unruh vacua [see Eqs.~(\ref{eq:N2}) and (\ref{eq:N4})], namely
\begin{equation}
\label{eq:N6}
\braket{N}^{\rm M}  \sim \frac{q^{2}d^{2}}{\textrm{min}[T_{1},T_{2}]^{2}} \, .
\end{equation}
This agrees with the estimate originally given in \cite{Belenchia_2018}. In particular, the decoherence effects do not grow with $T$. 

If we thermally populate the modes $\cev{R}_{\omega \ell}(r)$ in Minkowski spacetime at temperature $\mathscr T$, then the decoherence will be given by the same estimate as we previously obtained in Eq.~(\ref{eq:NHH}) for the decoherence effects of the incoming modes from infinity in Schwarzschild for the Hartle-Hawking vacuum, namely
\begin{equation}
\braket{N}_{\rm th.}^{\rm M} \sim \frac{q^{2}d^{2} {\mathscr  T}}{\textrm{min}[T_{1},T_{2}]} \, .
\end{equation}
In particular, the decoherence effects do not grow with $T$, despite the presence of the thermal bath. 

In a similar manner, in the gravitational case, for the Minkowski vacuum, we obtain
\begin{equation}
\label{eq:NGRMink2}
\braket{N}^{\rm M, GR} \sim \frac{m^{2}d^{4}}{\textrm{min}[T_{1},T_{2}]^{4}}
\end{equation}
in agreement with the original estimate of \cite{Belenchia_2018}. If Minkowski spacetime is populated with a thermal bath of gravitons at temperature $\mathscr T$, then we obtain the same estimate as in Eq.~(\ref{eq:NHH2}), namely
\begin{equation}
\label{eq:NHH3}
\braket{N}_{\rm th.}^{\rm M, GR}  \sim\frac{m^{2}d^{4} {\mathscr  T}}{\textrm{min}[T_{1},T_{2}]^{3}} .
\end{equation}
Again, the decoherence effects do not grow with $T$, despite the presence of a thermal bath of gravitons.

Finally, we point out that for a scalar field it is possible, in principle, to get decoherence in an inertial laboratory in Minkowski spacetime from ``soft radiation'' despite the absence of a horizon. In Minkowski spacetime, a memory effect and associated infrared divergences occur at null infinity for a massless field as a result of a permanent change in the field at order $1/r$. Since charge is conserved in electromagnetism, such $O(1/r)$ changes can occur in the electromagnetic case only via Lorentz boosting of the Coulomb fields of the charged particles. This generically occurs in scattering, since the outgoing charged particles generically have different momenta from the incoming particles. However, the protocol of Alice's experiment requires her to keep the components of her particle confined to her lab, which precludes changes in particle momenta lasting a long enough time $T$ to produce significant decoherence via ``soft radiation.'' This is in accord with what we have found above. Similarly, since mass is conserved in linearized gravity, there also are no significant ``soft radiation'' decoherence effects. However, for a scalar field, scalar charge need not be conserved, and a change in the scalar field at order $1/r$ can be achieved by simply changing the monopole moment of the source. Consequently, a source with a permanent change of scalar charge will radiate an infinite number of ``soft'' massless scalar particles in $\ell = 0$ modes. We can use this fact to obtain decoherence via soft radiation to null infinity in Minkowski spacetime in a manner previously suggested in \cite{Gralla:2023oya} as follows. 

Suppose that a massless scalar field $\op{\phi}$ exists in nature and Alice performs her experiment in an inertial laboratory in Minkowski spacetime with a particle with scalar charge. Suppose, further, that her protocol includes changing the charge of one of the components during separation and then restoring the charge during the recombination.\footnote{If the experiment is performed in the presence of a black hole or other gravitating body, such a change in scalar charge as determined at infinity automatically occurs from redshift effects if the components are separated in the radial direction \cite{Gralla:2023oya}.} The scalar analog of Eqs.~(\ref{eq:Nloc})~and~(\ref{eq:Ngrav}) is
\be
\braket{N} = \braket{\Omega|\left[\op{\phi}^{\textrm{in}}(j_{1}-j_{2}) \right]^{2}|\Omega}.
\label{Nscal}
\ee
The mode expansion of the two-point function of a scalar field in Schwarzschild is given in \cite{PhysRevD.21.2185}. It takes a form very similar to Eq.~(\ref{eq:Er2pt}) except that (i) the factor of $1/r^2{r'}^2$ is replaced by $1/rr'$ for the definition of scalar mode functions analogous to our definition of electromagnetic mode functions used in Eq.~(\ref{eq:Er2pt}) and (ii) the mode sum begins at $\ell = 0$ rather than $\ell = 1$. Only the incoming modes from infinity are relevant for Minkowski spacetime, and they again take the form Eq.~(\ref{inmink}). The $\ell=0$ modes contribute to Eq.~(\ref{Nscal}) an extra factor of $1/\omega^2$ relative to the $\ell = 1$ modes. For the case where the scalar field initially is in the Minkowski vacuum state $\ket{\Omega_{\rm M}}$, a calculation in direct parallel to Eq.~(\ref{eq:N2}) yields
\be
\braket{N}^{\rm M, S} \sim (\Delta q_{\rm S})^2 \ln\left(\frac{T}{\textrm{min}[T_{1},T_{2}]}\right)
\ee
where $\Delta q_{\rm S}$ denotes the scalar charge difference of the two components during their separation. This behavior is analogous to the decoherence occurring in the presence of a black hole for the Boulware vacuum [see Eqs.~(\ref{eq:Nright1}) and (\ref{eq:NHH2})]. If Minkowski spacetime is initially filled with a thermal bath of scalar particles at temperature $\mathscr T$, we obtain
\be
\braket{N}_{\rm th.}^{\rm M, S} \sim (\Delta q_{\rm S})^2 {\mathscr T} T
\ee
which is analogous to the decoherence in the presence of a black hole in the Unruh or Hartle-Hawking vacua.\footnote{{For a scalar field the similarity of the decoherence rate in a global thermal state in Minkowski spacetime, as compared to the decoherence due to a Killing horizon is related to the fact that {the restriction of the two-point function of the Minkowski vacuum to a uniformly accelerating world line is identical to the restriction of the two-point function of the global Minkowski thermal state at the Unruh temperature to an inertial world line. However,} for the electromagnetic and gravitational fields, no such equivalence holds \cite{PhysRevD.21.2137}, and as we have {seen,} these fields do not exhibit the analogous decoherence in a global thermal state.}} In both cases, the decoherence grows with $T$ due to the emission of soft radiation to infinity, and we thus see that such decoherence is possible, in principle, without the presence of a horizon.

\subsection{Decoherence in the Spacetime of a Static Star} 

{We now consider the decoherence effects arising in Alice's lab when we place it outside of a star rather than a black hole. In this subsection, we do not consider the decoherence effects that may arise from interactions with degrees of freedom of the matter composing the star, i.e., we are concerned only with the effects of replacing the black hole spacetime with a spacetime without a horizon. Decoherence effects due to interactions with matter will be considered in the next subsection.}

The metric outside of a static, spherical star is identical to the metric of a Schwarzschild black hole. If the electromagnetic field in the spacetime of a static star is initially in its ground state, then one might expect that if Alice performs her experiment outside of the star, she would get essentially the same results as she would have obtained by performing her experiment at the same radius in Schwarzschild spacetime with the electromagnetic field initially in the Boulware vacuum state.\footnote{In contrast to a static star, a body that collapses to a black hole produces the Unruh vacuum in its exterior, so that $\langle N \rangle $ grows linearly in time, as we have shown.} Similarly, if the electromagnetic field in the spacetime of the star is initially in a thermal state at temperature ${\mathscr T} = 1/8 \pi M$, one might expect that Alice would get essentially the same results as for a Schwarzschild black hole with the electromagnetic field initially in the Hartle-Hawking vacuum state. The purpose of this subsection is to explain why these expectations are not correct.

The key point is that the behavior of a quantum field in the spacetime of a star differs significantly from that of a quantum field around a black hole in that the white hole modes, $\vec{R}_{\omega \ell}(r)$, are absent. The complete absence of the white hole modes in the case of a star is very different from the modes being present but in their ground state, as occurs for the Boulware vacuum in Schwarzschild. The white hole modes in Schwarzschild represent additional degrees of freedom of the quantum field that are not present in the case of the star. It is these additional degrees of freedom---associated with the presence of a horizon---that are responsible for the decoherence effects that grow with $T$ in Alice's experiment. 

To see this explicitly, we note that in the spacetime of the star, the two-point function of the radial component of the electric field is modified from Eq.~(\ref{eq:Er2pt}) in that (i) the white hole modes, $\vec{R}_{\omega \ell}(r)$, are absent and (ii) the incoming modes from infinity, $\cev{R}_{\omega \ell}(r)$, are modified by the presence of the star. However, at very low frequencies, $\omega {\mathcal R} \ll 1$, where $\mathcal R$ denotes the radius of the star, the corrections to $\cev{R}_{\omega \ell}(r)$ are negligibly small. The ground state of the star satisfies $\cev{G}(\omega) = \Theta(\omega)$. It follows immediately that the decoherence in the spacetime of a star with the electromagnetic field initially in its ground state is the same as the decoherence in Schwarzschild due to the incoming modes from infinity in the Boulware or Unruh vacua [see Eq.~(\ref{eq:N4})], which, in turn, is the same as the decoherence in Minkowski spacetime in the Minkowski vacuum [see Eq.~(\ref{eq:N6})]. Thus, we obtain
\begin{equation}
\label{eq:N7}
\braket{N}^{\rm star}  \sim \frac{q^{2}d^{2}}{\textrm{min}[T_{1},T_{2}]^{2}} \, .
\end{equation}
Similarly, if the electromagnetic field around the star is in a thermal state at temperature $\mathscr T$, we obtain the same result as in Eq.~(\ref{eq:NHH2}), namely
\begin{equation}
\label{eq:NHH6}
\braket{N}_{\rm th.}^{\rm star} \sim \frac{q^{2}d^{2} {\mathscr  T}}{\textrm{min}[T_{1},T_{2}]} \, .
\end{equation}
In the gravitational case, we obtain results in agreement with Eqs.~(\ref{eq:NGRMink2}) and (\ref{eq:NHH3}), respectively.

In summary, the presence of a horizon is essential for the black hole decoherence effects. Similar effects do not occur in the spacetime of a static star.  

\subsection{Decoherence due to the Presence of a Body with Internal Degrees of Freedom}
\label{subsec:Newtent}
As we have just seen, in the electromagnetic and gravitational cases, decoherence due to emission of ``soft radiation'' does not occur in a static {asymptotically flat} spacetime without a horizon.\footnote{However, as discussed at the end of Sec.~\ref{minkowdec}, in the scalar case one can get decoherence due to emission of soft radiation to null infinity.} This can be understood as resulting from the absence of any ``white hole mode'' degrees of freedom associated with the horizon. However, if an actual material body is present, there will be additional degrees of freedom associated with the material body. These degrees of freedom can couple to the components of Alice's particle via ordinary Coulombic (or, in the gravitational case, Newtonian) interactions. If there is suitable dissipation in the material body system, this can result in the decoherence of Alice's particle. Indeed, ordinary environmental decoherence is exactly of this nature. In this subsection, we will consider whether the decoherence of Alice's particle resulting from Coulombic/Newtonian interactions with a material body can mimic the decoherence obtained for the case of a black hole.

As we have seen in Sec. \ref{locsch} above, in the electromagnetic case the dominant contribution to decoherence of Alice's particle near a Schwarzschild black hole in the Unruh vacuum comes from the $\ell = 1$ white hole modes at very low frequencies. Very near the horizon of the black hole, these modes correspond to radiation and they represent genuine additional degrees of freedom of the electromagnetic field. Nevertheless, we saw at the end of Sec. \ref{locsch} that in Alice's lab, these modes look just like the exterior dipole field of an ordinary body, with a fluctuating electric dipole moment given by Eq.~(\ref{fluctdip}). Thus, if we have a material body with the property that its ordinary thermal fluctuations cause its electric dipole moment at very low frequencies $\omega$ to fluctuate in accord with Eq.~(\ref{fluctdip}), then that material body should mimic the decoherence effects of a black hole. Similarly, in the gravitational case, a material body will mimic the decoherence effects of a black hole if ordinary thermal fluctuations cause its mass quadrupole moment at very low frequencies $\omega$ to fluctuate in accord with Eq.~(\ref{fluctquad}).

The issue of whether an ordinary material body can mimic a black hole of the same temperature in this manner has very recently been investigated by Biggs and Maldacena \cite{Biggs:2024dgp}. They have shown that in the electromagnetic case, there are no difficulties in constructing a physically reasonable matter model that mimics the ``soft radiation'' decoherence effects of a black hole. However, in the gravitational case, the mimicking of black hole decoherence effects by an ordinary body of the same physical size {and temperature} as the black hole appears to require extraordinary properties of the matter. The underlying difficulty is the weakness of the coupling of matter to gravity. In order to produce a fluctuating quadrupole moment of the required size Eq.~(\ref{fluctquad}), it seems possible that the body would need to have a mass comparable to that of a black hole as well as extremely large dissipation. This issue appears worthy of further investigation. 
\begin{acknowledgments}
We thank Anna Biggs, Simon Carot-Huot, Yanbei Chen, Juan Maldacena, and Jordan Wilson-Gerow for helpful discussions. D.L.D. acknowledges support as a Fannie and John Hertz Foundation Fellow holding the Barbara Ann Canavan Fellowship and as an Eckhardt Graduate Scholar at the University of Chicago. This research was supported in part by NSF Grants No. 21-05878 { and 24-03584} and Templeton Foundation Grant No. 62845 to the University of Chicago and by the Princeton Gravity Initiative at Princeton University.
\goodbreak
\end{acknowledgments}
\bibliography{localDecoherence.bib}
\end{document}